**Path-integral evolution of multivariate systems with moderate noise**

Lester Ingber

*Lester Ingber Research, P.O. Box 857, McLean, VA 22101*

ingber@alumni.caltech.edu

A non Monte Carlo path-integral algorithm that is particularly adept at handling nonlinear Lagrangians is extended to multivariate systems. This algorithm is particularly accurate for systems with moderate noise.

PATHINT is a non Monte Carlo histogram C computer code developed to evolve an $n$-dimensional system (subject to machine constraints) based on a generalization of an algorithm demonstrated by Wehner and Wolfer to be extremely robust for nonlinear Lagrangians with moderate noise [1-3]. PATHINT was recently used in a neuroscience study [4], where it was observed how difficult it can be even in two dimensions to get good resolution because of CPU constraints on Sun SPARC 2 and 10MP machines. Here it appears that the resolution is quite satisfactory on these machines.

The system selected for this paper to illustrate the use of PATHINT is the classical analog of a quantum system. "Quantum chaos" was a term used to describe the observation of chaos in the classical trajectories of the Hamilton's equations of motion,

$$\dot{p} = -\frac{\partial H}{\partial q},$$

$$\dot{q} = \frac{\partial H}{\partial p}, \tag{1}$$

where a Hamiltonian of the form

$$H = \frac{1}{2} p^2 + \hat{\Phi}(q),$$

$$\hat{\Phi} = \frac{1}{2} (\nabla \Phi)^2 - \frac{1}{2} D \nabla^2 \Phi, \tag{2}$$

with a coefficient $D$ was considered in a $q$ dimension $\geq 2$, representing double this dimension in the corresponding phase space of $q$ and $p$ [5].

In ref. [5] a potential $\Phi$ was considered,

$$\Phi = 2x^4 + \frac{3}{5} y^4 + \varepsilon xy(x - y)^2, \tag{3}$$

in a classical Fokker-Planck system

$$\partial P_t = \nabla . (P \nabla \Phi) + \frac{D}{2} \nabla^2 P. \tag{4}$$

A slight change of notation is used in this paper. They posed the question as to just what properties such classical systems might possess?

This paper does not at all deal with the quantum system described above, but it does deal in detail with the associated classical system, and it does give some answers to the above posed question. This study should be considered as illustrating a particular numerical approach that promises to be quite useful in studying the evolution of such classical systems. The results obtained can be considered as "experimental" data on the exact region of such classical transformations. The results here are negative with respect to any unusual or interesting activity in the parameter region observed in the quantum mechanical calculation. This should at least help other investigators who might tend to focus on this region in the classical system, based on the results obtained for the associated quantum system.

This paper computes the path integral of the classical system in terms of its Lagrangian $L$.

$$P[q_t|q_{t_0}]dq(t) = \int \cdots \int Dq \exp\left(-\min \int_{t_0}^{t} dt' L\right) \delta(q(t_0) = q_0)\delta(q(t) = q_t) ,$$

$$Dq = \lim_{u \to \infty} \prod_{\rho=1}^{u+1} g^{1/2} \prod_i (2\pi\Delta t)^{-1/2} dq_\rho^i ,$$

$$L(\dot{q}^i, q^i, t) = \frac{1}{2}(\dot{q}^i - g^i)g_{ii'}(\dot{q}^{i'} - g^{i'}) ,$$

$$g_{ii'} = (g^{ii'})^{-1} ,$$

$$g = \det(g_{ii'}) . \tag{5}$$

Here the diagonal diffusion terms are $g^{xx} = g^{yy} = D$ and the drift terms are $g^i = -\partial \Phi / \partial q^i$. If the diffusions terms are not constant, then there are additional terms [6].

The histogram procedure recognizes that the distribution can be numerically approximated to a high degree of accuracy by sums of rectangles of height $P_i$ and width $\Delta q^i$ at points $q^i$. For convenience, just consider a one-dimensional system. The above path-integral representation can be rewritten, for each of its intermediate integrals, as

$$P(x; t + \Delta t) = \int dx' [g^{1/2} (2\pi \Delta t)^{-1/2} \exp(-L \Delta t)] P(x'; t)$$

$$= \int dx' G(x, x'; \Delta t) P(x'; t) ,$$

$$P(x; t) = \sum_{i=1}^{N} \pi(x - x^i) P_i(t) ,$$

$$\pi(x - x^i) = \begin{cases} 1, & (x^i - \frac{1}{2} \Delta x^{i-1}) \leq x \leq (x^i + \frac{1}{2} \Delta x^i) , \\ 0, & \text{otherwise} . \end{cases} \quad (6)$$

This yields

$$P_i(t + \Delta t) = T_{ij}(\Delta t) P_j(t) ,$$

$$T_{ij}(\Delta t) = \frac{2}{\Delta x^{i-1} + \Delta x^i} \int_{x^i - \Delta x^{i-1}/2}^{x^i + \Delta x^i/2} dx \int_{x^j - \Delta x^{j-1}/2}^{x^j + \Delta x^j/2} dx' G(x, x'; \Delta t) . \quad (7)$$

$T_{ij}$ is a banded matrix representing the Gaussian nature of the short-time probability centered about the (possibly time-dependent) drift.

This histogram procedure was extended to two dimensions using a matrix $T_{ijkl}$ [7]. Explicit dependence of $L$ on time $t$ also can be included without complications. Care must be used in developing the mesh in $\Delta q^i$, which is strongly dependent on the diagonal elements of the diffusion matrix, e.g.,

$$\Delta q^i \approx (\Delta t g^{ii})^{1/2} . \quad (8)$$

This constrains the dependence of the covariance of each variable to be a (nonlinear) function of that variable in order to present a straightforward rectangular underlying mesh.

Since integration is inherently a smoothing process [8], fitting the data with the short-time probability distribution, effectively using an integral over this epoch, permits the use of coarser meshes than the corresponding stochastic differential equation. For example, the coarser resolution is appropriate, as typically required, for a numerical solution of the time-dependent path integral. By considering the contributions to the first and second moments, conditions on the time and variable meshes can be derived [1]. The time slice essentially is determined by $\Delta t \leq \bar{L}^{-1}$, where $\bar{L}$ is the uniform Lagrangian, respecting ranges giving the most important contributions to the probability distribution $P$. Thus $\Delta t$ is roughly measured by the diffusion divided by the square of the drift.

Such calculations are useful in many disciplines, e.g., some financial instruments [8,9]. Monte Carlo algorithms for path integrals are well known to have extreme difficulty in evolving nonlinear systems with multiple optima [10], but this algorithm does very well on such systems. The PATHINT code was tested against the test problems given in previous one-dimensional systems [1,2], and it was established that the method of images for both Dirichlet and Neumann boundary conditions is as accurate as the boundary element methods for the systems investigated. Two-dimensional runs were tested by using cross products of one-dimensional examples whose analytic solutions are known.

Attempts were made to process the same system considered for the quantum case [5]. Therefore, the diffusion was taken to be $D = 0.2$. Since they selected an harmonic oscillator basis for their eigenvalue study, is was assumed that natural boundary conditions are appropriate for this study, and ranges of $x$ and $y$ were tested to ensure that this was reasonable. A band of three units on each side of the short-time distribution was sufficient for these runs. For $\varepsilon \leq 5$, the range of $x$ was taken to be $\pm 3$ and the range of $y$ was taken to be $\pm 6$.

A mesh of $\Delta t = 0.1$ was reasonable to calculate the evolution of this system for $0.1 \leq \varepsilon \leq 0.5$. The quantum study observed chaos at $\varepsilon \geq 0.14$, but the classical system appears to be very stable with a single peak in its probability density up through $\varepsilon = 0.5$. The time mesh was tested by performing several calculations at time meshes of $\Delta t = 0.01$ on a Sun SPARC 10MP. All other calculations reported here were performed on a Sun SPARC 2.

Figures 1(a) and 1(b) show the evolution of the distribution for $\varepsilon = 0.1$, $t = 10$ and 100, i.e., after 100 and 1000 foldings of the path integral, respectively. The distribution starts at single peaks at the imposed initial condition, a $\delta$ function at the origin, i.e., $x = -0.0302$ and $y = -0.0603$ with this mesh, and swell out their stable structures within $t < 0.03$. Note the stability over the duration of the calculation. A similar stability was noted for $\varepsilon < 0.55$, as illustrated in Figures 2(a) and 2(b) for $\varepsilon = 0.3$ and 0.5 after 1000 foldings.

For values of $\varepsilon > 0.5$, a mesh of $\Delta t = 0.05$ was used to investigate the onset of instabilities which were noted for higher $\varepsilon$ with coarser meshes. To two significant figures, this first occurs at $\varepsilon = 0.55$. As $\varepsilon$ increases, so does the peak spread quite early along a diagonal in the evolving distribution. As the time of the calculation increased, there was concern that the boundaries were being approached. Therefore, for these runs, the range of $x$ was increased to $\pm 5$ and the range of $y$ was increased to $\pm 10$. Figure 3 illustrates the evolution at time $t = 5.0$ (100 foldings) and $t = 100$ (2000 foldings) with $\varepsilon = 0.55$. Figure 4 illustrates the early structure for $\varepsilon = 0.6$ at $t = 15$ after 300 foldings of $\Delta t = 0.05$.

The PATHINT algorithm utilized here will be used to explore the time evolution of other Fokker-Planck systems in the presence of moderate noise. As mentioned above, such problems arise in many systems ranging from neocortical interactions to financial markets. Also, we can now accurately examine long-time correlations of chaotic models as multiplicative noise is increased to moderate and strong levels; many chaotic models do not include such levels of noise as is found in the systems they are attempting to model. A project is now underway under an award of Cray computer time from the Pittsburgh Supercomputing Center through the National Science Foundation (NSF), the Parallelizing ASA and PATHINT Project (PAPP) further developing this code for large systems [11].

## FIGURE CAPTIONS

FIG. 1. Probability density for $\varepsilon = 0.1$, for $\Delta t = 0.1$ (a) after 100 foldings ($t = 10$) and (b) after 1000 foldings ($t = 100$).

FIG. 2. Probability density after 1000 foldings of $\Delta t = 0.1$ ($t = 100$) for $\varepsilon$ set to (a) 0.3 and (b) 0.5.

FIG. 3. Probability density for $\varepsilon$ set to 0.55 using $\Delta t = 0.05$, (a) at $t = 5$ after 100 foldings, and (b) at $t = 100$ after 2000 foldings. As discussed in the text, the ranges of $x$ and $y$ for $\varepsilon > 0.5$ were increased.

FIG. 4. Probability density at $t = 15$, after 300 foldings of $\Delta t = 0.05$, for $\varepsilon = 0.6$.

**Figure 1a**

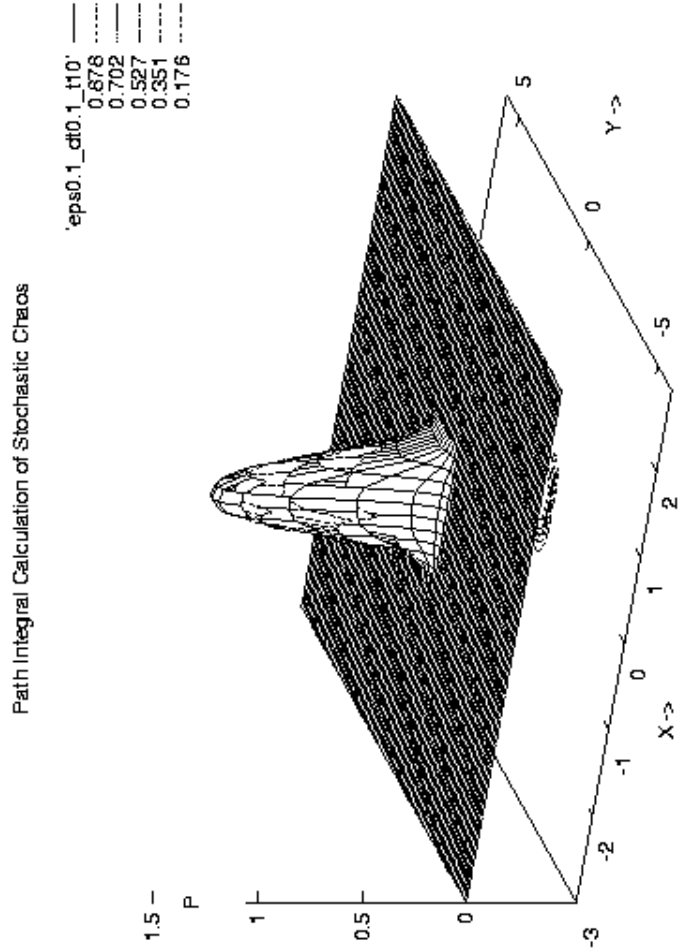

**Figure 1b**

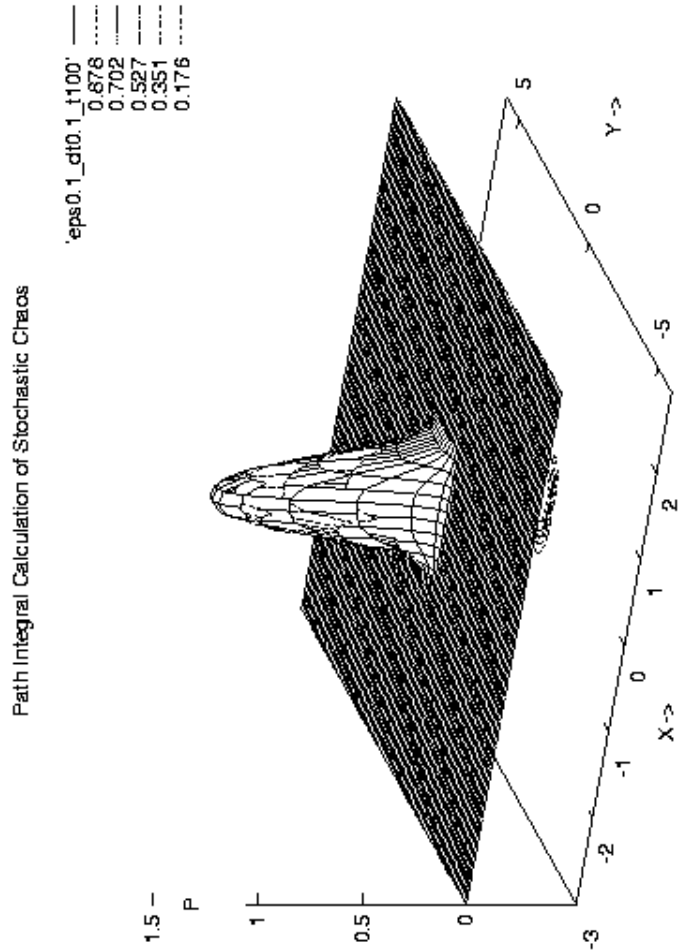

**Figure 2a**

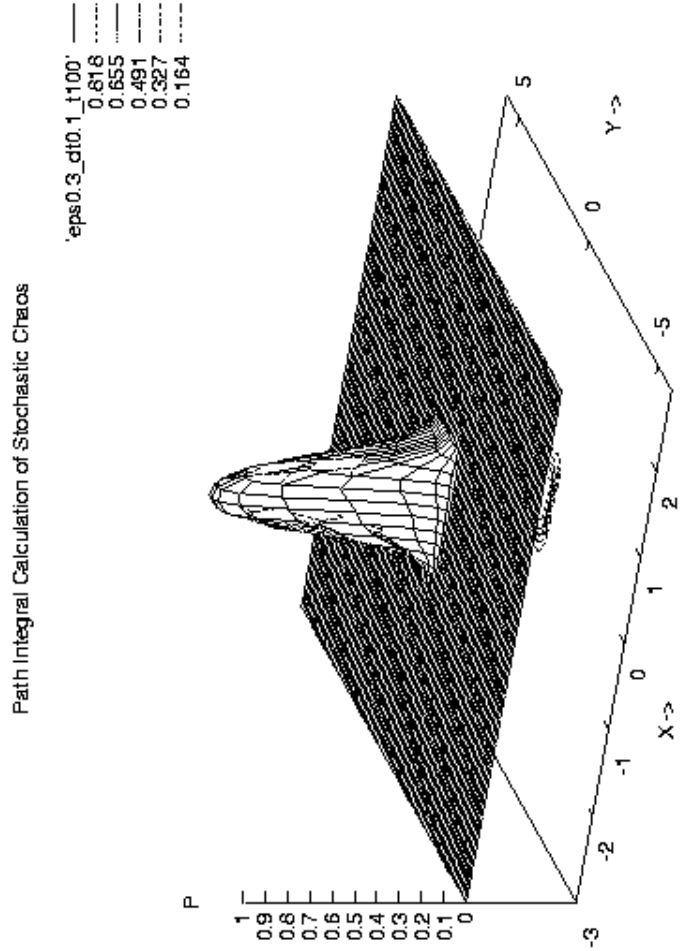

**Figure 2b**

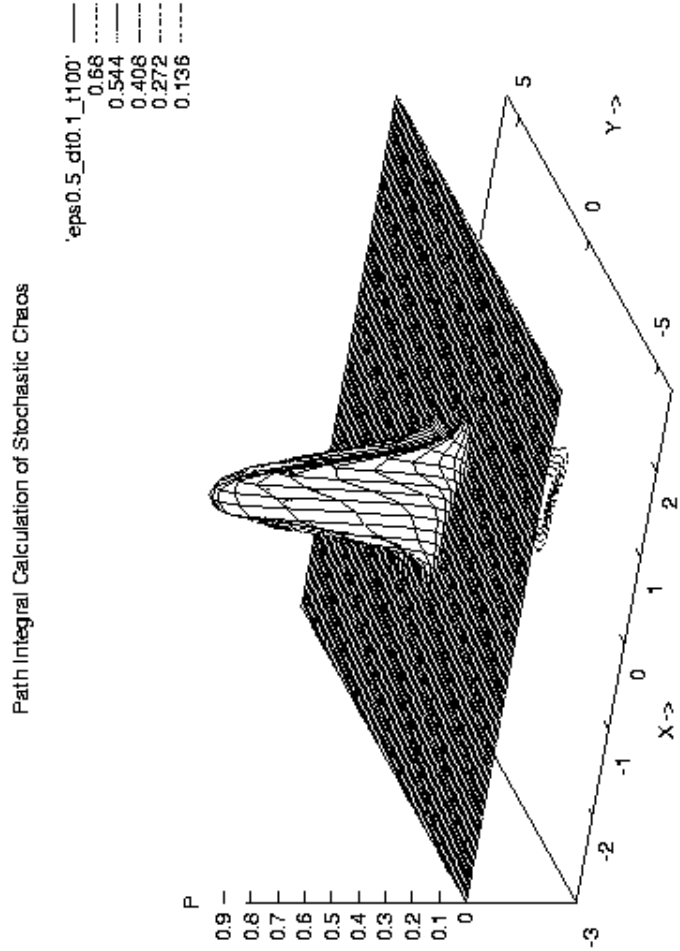

**Figure 3a**

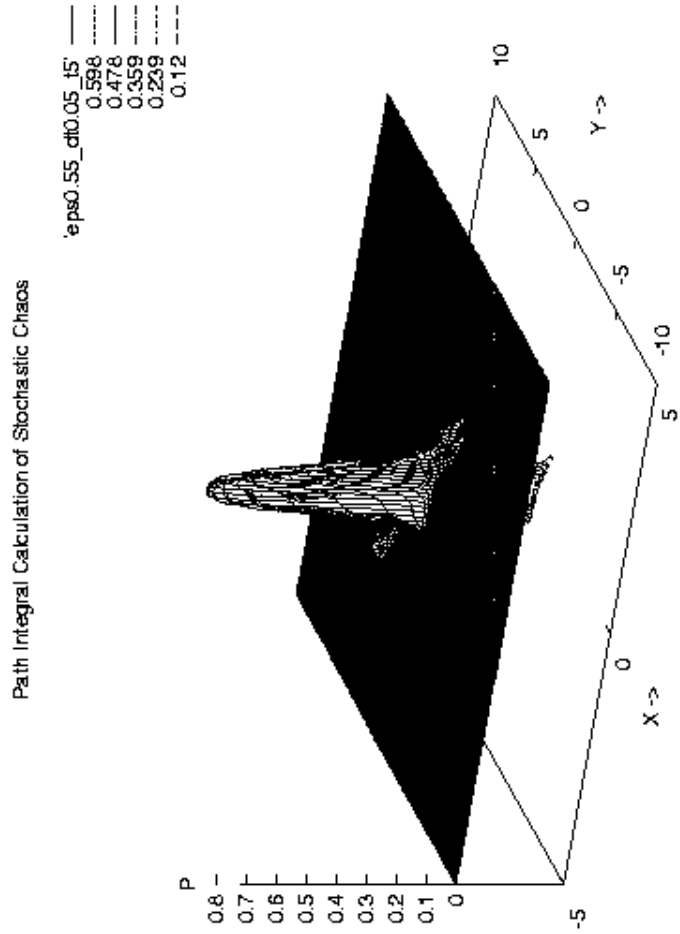

**Figure 3b**

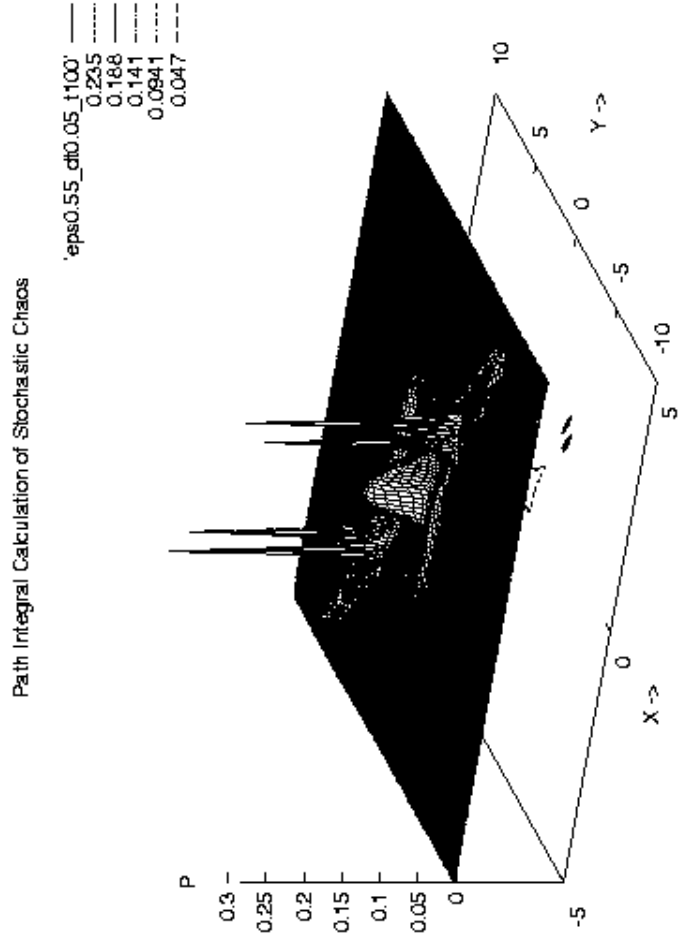

**Figure 4**

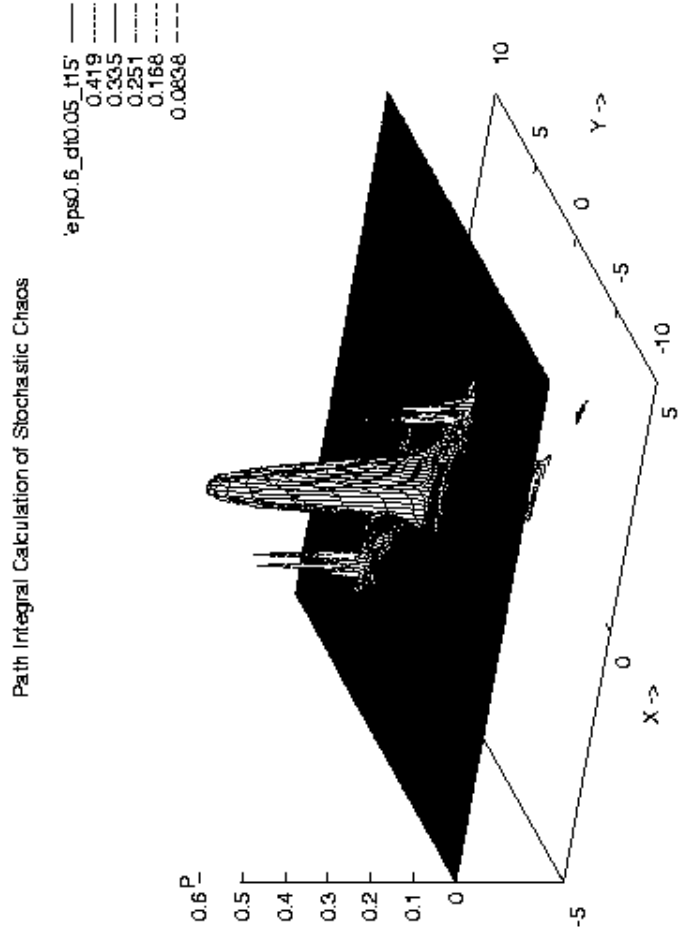